\documentclass[12pt]{article}

\usepackage[a4paper,left=30mm,right=20mm,top=20mm,bottom=20mm]{geometry}
\usepackage{graphics}
\usepackage{amsmath}
\usepackage{epsfig}
\usepackage{cite}
\usepackage{xcolor}

\newcommand{\bea}{\begin{eqnarray}}
\newcommand{\eea}{\end{eqnarray}}
\newcommand{\be}{\begin{equation}}
\newcommand{\ee}{\end{equation}}

\newcommand{\ar}{a_s}

\title{On Fractional Analytic QCD}

\author{A.V.~Kotikov$^{1}$, I.A.~Zemlyakov$^{1,2}$}

\begin{document}

\maketitle

\begin{center}
 
{\it $^{1}$Joint Institute for Nuclear Research, 141980, Dubna, Moscow region, Russia}\\
{\it  $^2$Tomsk State University,
  634010 Tomsk,
  Russia}

\end{center}

\vspace{0.5cm}

\begin{center}

{\bf Abstract }

\end{center}

We present a brief  overview of fractional analytic QCD.\\

\noindent

\label{sec:intro}
\section{Introduction}

The strong coupling constant (couplant) $\alpha_s(Q^2)$ obeys  the renormalization group equation
\be
L\equiv \ln\frac{Q^2}{\Lambda^2} = \int^{\overline{a}_s(Q^2)} \, \frac{da}{\beta(a)},~~ \overline{a}_s(Q^2)=\frac{\alpha_s(Q^2)}{4\pi}\,,~~\ar(Q^2)=
\beta_0\,\overline{a}_s(Q^2)\,,
\label{RenGro}
\ee 
with some boundary condition and the QCD $\beta$-function:
\be
\beta(\ar)~=~
-\beta_0  \overline{a}_s^{2} \, \Bigl(1+\sum_{i=1} b_i \ar^i \Bigr),~~
 \beta_0=11-\frac{2f}{3},~~\beta_1=102-\frac{38f}{3},
\label{beta}
\ee
for $f$ active quark flavors. Really now the first fifth coefficients
are exactly known \cite{Baikov:2016tgj}.
In our present consideration we will need only $i=0$ and $i=1$.

So, already at leading order (LO), where $\ar(Q^2)=\ar^{(1)}(Q^2)$, we have from Eq. (\ref{RenGro})
\be
\ar^{(1)}(Q^2) = \frac{1}{L}\, ,
\label{asLO}
\ee
i.e. $\ar^{(1)}(Q^2)$ does contain a pole at $Q^2=\Lambda^2$.

In \cite{ShS}, an efficient approach was developed to eliminate the Landau singularity. It is based on the dispersion relation, which
connects the new analytic couplant $A_{\rm MA}(Q^2)$ with the spectral function $r_{\rm pt}(s)$
obtained in the PT framework.
In LO this gives
    \be
A^{(1)}_{\rm MA}(Q^2) 
= \frac{1}{\pi} \int_{0}^{+\infty} \, 
\frac{ d s }{(s + t)} \, r^{(1)}_{\rm pt}(s),~~ r^{(1)}_{\rm pt}(s)= {\rm Im} \; a_s^{(1)}(-s - i \epsilon) \,.
\label{disp_MA_LO}
\ee

This approach is usually
called  {\it Minimal Approach} (MA) (see, e.g., \cite{Cvetic:2008bn})
or  {\it Analytical Perturbation Theory} (APT) \cite{ShS}.

Further APT development  is the so-called fractional APT (FAPT) \cite{BMS1}, which extends the principles  of construction 
described above to PT series starting with non-integer degrees of the couplant. Within  the framework of QFT, such series arise
for quantities having
non-zero anomalous dimensions.

In this short paper we give an overview of the basic properties of MA couplants, obtained in \cite{Kotikov:2022sos}
using the so-called $1/L$-expansion. Note that for an ordinary couplant, this expansion is only applicable for large $Q^2$ values,
i.e. for $Q^2>>\Lambda^2$.
However, as shown in  \cite{Kotikov:2022sos,KoZe23},  the situation is  completely  different in the case of analytic couplants:
this $1/L$-expansion is applicable for all  argument values. This is due to the fact that non-leading corrections to the expansion
disappear not only at $Q^2 \to \infty$, but also at $Q^2 \to 0$,
which leads only to nonzero (small) corrections in the region $Q^2 \sim \Lambda^2$. 

Below we consider the representations for the MA couplants and their (fractional) derivatives obtained in
\cite{Kotikov:2022sos,KoZe23} in principle, in any PT order. However, in order to avoid cumbersome formulas, but at the same time
to show the main features of the approach, we limit ourselves to considering only the first two PT orders.

\section{Strong couplant}
\label{strong}

As shown in the Introduction, $a_s(Q^2)$ obeys the renormalized group equation (\ref{RenGro}).
When $Q^2>>\Lambda^2$, Eq. (\ref{RenGro})  can be solved by iterations in the form of a $1/L$-expansion,
which can be represented in the following compact form
\be
a^{(1)}_{s,0}(Q^2) = \frac{1}{L_0},~~
a^{(2)}_{s,1}(Q^2) = 
a^{(1)}_{s,1}(Q^2) +
\delta^{(2)}_{s,1}(Q^2)
\label{as}
\ee
where
\be
\delta^{(2)}_{s,k}(Q^2) = - \frac{b_1\ln L_k}{L_k^2} ,~~
L_k=\ln t_k,~~t_k=\frac{1}{z_k}=\frac{Q^2}{\Lambda_k^2}\,.
\label{L}
\ee

So, in any PT order, the couplant $\ar(Q^2)$ contains its own dimensional transmutation parameter $\Lambda$, which is related to the
normalization $\alpha_s(M_Z^2)$,
where $\alpha_s(M_Z)=0.1176$ in  PDG20 \cite{PDG20} (see also \cite{Enterria}).

{\bf $f$-dependence of the couplant $\ar(Q^2)$.}~~
The coefficients $\beta_i$ in (\ref{beta}) depend on the number $f$ of active quarks that change the couplant $\ar(Q^2)$ at
threshold values $Q^2_f \sim m^2_f$, when some additional quark comes into play at $Q^2 > Q^2_f$.
Thus, the couplant $a_s$ depends on $f$, and this $f$-dependence can be taken into account in $\Lambda$, i.e. it is
$\Lambda^f$ contributes to the above Eqs. (\ref{RenGro}) and (\ref{as}).

The relationship between $\Lambda_{i}^{f}$ and $\Lambda_{i}^{f-1}$ is known up to the four-loop order \cite{Chetyrkin:2005ia} in
the $\overline{MS}$ scheme.
%
Here we will not consider the $f$-dependence of $\Lambda_{i}^{f}$,
since we are mainly considering the range of small $Q^2$ values and therefore use
$\Lambda_{i}^{f=3}$ (see, e.g., \cite{Chen:2021tjz}):
\be
\Lambda_0^{f=3}=142~~ \mbox{MeV},~~\Lambda_1^{f=3}=367~~ \mbox{MeV}
\,.
\label{Lambdas}
\ee

\section{Fractional derivatives}

Following \cite{Cvetic:2006mk,Cvetic:2006gc},
we introduce the derivatives (in the $(i)$-order of of PT)
\be
\tilde{a}^{(i)}_{n+1}(Q^2)=\frac{(-1)^n}{n!} \, \frac{d^n a^{(i)}_s(Q^2)}{(dL)^n} \, ,
\label{tan+1}
\ee
which are very convenient in the case of the analytic QCD (see, e.g., \cite{Kotikov:2022JETP}).

The series of derivatives $\tilde{a}_{n}(Q^2)$ can successfully replace the corresponding series of $\ar$-degrees. Indeed, each
derivative reduces the $\ar$ degree, but is accompanied by an additional $\beta$-function $\sim \ar^2$.
Thus, each application of a derivative yields an additional $\ar$, and thus it is really possible to use series of derivatives
instead of series of $\ar$-powers.

In LO, the series of derivatives $\tilde{a}_{n}(Q^2)$ are exactly the same as $\ar^{n}$. Beyond LO, the relation between
$\tilde{a}_{n}(Q^2)$ and $\ar^{n}$ was established in \cite{Cvetic:2006gc,Cvetic:2010di}
and extended to fractional cases, where $n \to$ a non-integer $\nu $
in \cite{GCAK}.

Now consider the $1/L$-expansion of $\tilde{a}^{(k)}_{\nu}(Q^2)$ $(k=0,1)$ at LO and next-to-leading (NLO) approximations
\be
\tilde{a}^{(1)}_{\nu,0}(Q^2)={\bigl(a^{(1)}_{s,0}(Q^2)\bigr)}^{\nu} = \frac{1}{L_0^{\nu}},~
\tilde{a}^{(2)}_{\nu,1}(Q^2)=\tilde{a}^{(1)}_{\nu,1}(Q^2) +
\nu\, \tilde{\delta}^{(2)}_{\nu,1}(Q^2),~~
\label{tdmp1N}
\ee
where
\be
\tilde{\delta}^{(2)}_{\nu,1}(Q^2)=
\hat{R}_1 \, \frac{1}{L_i^{\nu+1}}= \Bigl[\hat{Z}_1(\nu)+ \ln L_i\Bigr]\, \frac{1}{L_i^{\nu+1}},~~
\hat{R}_1=b_1 \Bigl[\hat{Z}_1(\nu)+ \frac{d}{d\nu}\Bigr],~~
\label{hR_i}
\ee
with $\hat{Z}_1(\nu)=\Psi(\nu+1)+\gamma_{E}-1$, where $\Psi(\nu+1)$ and $\gamma_{E}$ are Euler's constant and $\Psi$-function,
respectively.

Representation (\ref{tdmp1N}) of the $\tilde{\delta}^{(2)}_{\nu,1}(Q^2)$ correction in the form of the $\hat{R}_1$-operator is very
important and allows us to similarly represent high-order results for the ($1/L$-expansion) of analytic couplants.

\section{MA coupling}

We  first show the LO results, and then the NLO ones
following our results (\ref{tdmp1N}).

{\bf LO.}~~
The LO MA couplant $A^{(1)}_{{\rm MA},\nu,0}$
has the following form  \cite{BMS1}
\be
A^{(1)}_{{\rm MA},\nu,0}(Q^2) = {\left( a^{(1)}_{\nu,0}(Q^2)\right)}^{\nu} - \frac{{\rm Li}_{1-\nu}(z_0)}{\Gamma(\nu)}
\equiv \frac{1}{L_0^{\nu}}-\Delta^{(1)}_{\nu,0}\,,
\label{tAMAnu}
\ee
where
\be
   {\rm Li}_{\nu}(z)=\sum_{m=1}^{\infty} \, \frac{z^m}{m^{\nu}}=  \frac{z}{\Gamma(\nu)} \int_0^{\infty} 
\frac{ dt \; t^{\nu -1} }{(e^t - z)}
   \label{Linu}
\ee
is the Polylogarithm.
For $\nu=1$ we recover the famous Shirkov-Solovtsov results  \cite{ShS}:
\be
A^{(1)}_{\rm MA,0}(Q^2) \equiv A^{(1)}_{\rm MA,\nu=1,0}(Q^2)
=\frac{1}{L_0}- \frac{z_0}{1-z_0}\,,
\label{tAM1}
\ee
which
can be taken directly for the integral forms (\ref{disp_MA_LO}).

{\bf NLO.}~~ By analogy with ordinary couplant,
using the results (\ref{tdmp1N})
we have for MA analytic couplantt $\tilde{A}^{(i+1)}_{{\rm MA},\nu,i}$
the following expressions:
\be
\tilde{A}^{(2)}_{{\rm MA},\nu,1}(Q^2) = \tilde{A}^{(1)}_{{\rm MA},\nu,1}(Q^2) +
\nu \, \tilde{\delta}^{(2)}_{{\rm MA},\nu,i}(Q^2),
\label{tAiman}
\ee
where  $\tilde{A}^{(1)}_{{\rm MA},\nu,i}$
is given in Eq.  (\ref{tAMAnu})
and
\be
\tilde{\delta}^{(2)}_{{\rm MA},\nu,1}(Q^2)= \tilde{\delta}^{(2)}_{\nu,1}(Q^2) -\hat{R}_1
\left( \frac{{\rm Li}_{-\nu}(z_1)}{\Gamma(\nu+1)}\right)=\tilde{\delta}^{(2)}_{\nu,1}(Q^2) - \Delta^{(2)}_{\nu,1}(z_1),
\label{tdAman}
\ee
with $\overline{\gamma}_{\rm E}=\gamma_{\rm E}-1$,
\be
\Delta^{(2)}_{\nu,1}(z)
=b_1\Bigl[\overline{\gamma}_{\rm E}
  {\rm Li}_{-\nu}(z)+{\rm Li}_{-\nu,1}(z)\Bigr],~~
 {\rm Li}_{\nu,1}(z)= \sum_{m=1}  \,\frac{z^m \,\ln m}{m^{\nu}},~~
 {\rm Li}_{-1}(z)= \frac{z}{(1-z)^2}
    \, .
\label{Lii.1}
\ee
%
and $\tilde{\delta}^{(2)}_{\nu,1}(Q^2)$ and $\hat{R}_1$
are given in Eqs. (\ref{tdmp1N}) and (\ref{L}), respectively.

The
results for the MA analytic couplant $\tilde{A}^{(i+1)}_{{\rm MA},\nu,i}$
can be found if
$\nu=1$.

\begin{figure}[!htb]
\centering
\includegraphics[width=0.58\textwidth]{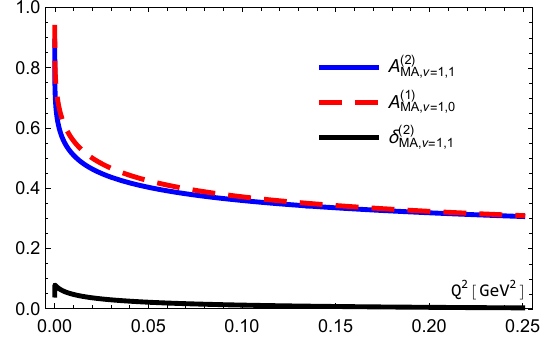}
    \caption{\label{fig:Adelta2}
      The results for $A^{(1)}_{\rm MA,\nu=1,0}(Q^2)$, $A^{(2)}_{\rm MA,\nu=1,1}(Q^2)$ and $\delta^{(2)}_{\rm MA,\nu=1,1}(Q^2)$.
    }
\end{figure}

On Fig. \ref{fig:Adelta2} we see that $A^{(i+1)}_{\rm MA,i}(Q^2)$ 
are very close to each other for $i=0$ and $i=1$.
The differences $\delta^{(2)}_{\rm MA,\nu=1,1}(Q^2)$
between the L0 and NLO results are nonzero only for $Q^2 \sim \Lambda^2$.

\section{MA coupling. Another form}

The results (\ref{tAMAnu}) and (\ref{tAiman}) for MA  couplants are very convenient in the range of large and small values of $Q^2$.
For $Q^2 \sim \Lambda_i^2$, both parts,
the standard couplant and the additional term $\delta^{(i+1)}_{\rm MA,\nu,i}(Q^2)$, have singularities that cancel out in sum.
Thus, numerical applications of these results may not be so simple, requiring, for example, some sub-expansions for each part
in the neighborhood of the point $Q^2 =\Lambda_i^2$ .
Therefore, here we propose another form that is very useful for $Q^2 \sim \Lambda_i^2$ and can be used for any value of $Q^2$ as well, except for the ranges of very large and very small $Q^ 2$ values.

{\bf LO.}~~
The LO MA coupling $A^{(1)}_{{\rm MA},\nu}(Q^2)$ \cite{ShS}
has also the another form  \cite{BMS1}
\be
A^{(1)}_{{\rm MA},\nu}(Q^2) =
\frac{(-1)}{\Gamma(\nu)} \, \sum_{r=0}^{\infty} \zeta(1-\nu-r) \, \frac{(-L)^r}{r!}~ (L<2\pi),~~\zeta(\nu)
     =\sum_{m=1}^{\infty} \, \frac{1}{m^{\nu}}
\label{tAMAnuNew}
\ee
where $\zeta(\nu)$ are Euler $\zeta$-functions.

The result (\ref{tAMAnuNew}) was obtained in Ref. \cite{BMS1} using properties of the Lerch function, which can be considered as a
generalization of Polylogarithms (\ref{Linu}).

For $\nu=1$ we have
\be
A^{(1)}_{\rm MA}(L) = - \, \sum_{r=0}^{\infty} \zeta(-r) \, \frac{(-L)^r}{r!},~~ \zeta(-r) = (-1)^{r} \, \frac{B_{r+1}}{r+1}
\label{tAMA1New}
\ee
where $B_{r+1}$ are Bernoulli numbers.
Using their properties, we have for even $r=2m$ and for odd  $r=1+2l$ values
\be
\zeta(-2m) = -\frac{\delta^0_m}{2},~~\zeta(-(1+2l))= - \frac{B_{2(l+1)}}{2(l+1)} \,,
\label{ze_r1}
\ee
where $\delta^0_m$ is the Kronecker symbol.
Thus, for $A^{(1)}_{\rm MA}(Q^2)$  we have ($s=l+1$)
\be
A^{(1)}_{\rm MA}(Q^2) = \frac{1}{2} \, \left(1
+ \, \sum_{l=0}^{\infty} \frac{B_{2(l+1)}}{l+1} \, \frac{(-L)^{2l+1}}{(2l+1)!}\right)=\frac{1}{2} \, \left(1
+ \, \sum_{s=1}^{\infty} \frac{B_{2s}}{s} \, \frac{(-L)^{2s-1}}{(2s-1)!}\right) \,.
\label{tAMA1New.1}
\ee

{\bf NLO.}~~
Now we consider the derivatives of MA coupling constant, i.e. $\tilde{A}^{(1)}_{{\rm MA},\nu}$, shown
in Eq. (\ref{tAiman}), i.e.
\be
\tilde{A}^{(2)}_{{\rm MA},\nu,1}(Q^2) = \tilde{A}^{(1)}_{{\rm MA},\nu,1}(Q^2) +
\nu\,\tilde{\delta}^{(2)}_{{\rm MA},\nu,1}(Q^2)\,,~~
\tilde{\delta}^{(2)}_{{\rm MA},\nu,1}(Q^2)= \hat{R}_1   \, A^{(1)}_{{\rm MA},\nu+1,1} \, ,
\label{tdAmanNew}
\ee
where operators $\hat{R}_1$ are given above in (\ref{L}).
After some calculations we have
\be
\tilde{\delta}^{(2)}_{{\rm MA},\nu,1}(Q^2)=
\frac{(-1)}{\Gamma(\nu+1)} \, \sum_{r=0}^{\infty} \tilde{R}_1(\nu+r) \, \frac{(-L_k)^r}{r!}
\label{tdAmanNew}
\ee
where
\be
\tilde{R}_1(\nu+r)=b_1\Bigl[\overline{\gamma}_{\rm E}
  \zeta(-\nu-r)+\zeta_1(-\nu-r)\Bigr],~~
\zeta_k(\nu)=
\sum_{m=1}^{\infty} \, \frac{\ln^k m}{m^{\nu}}
\, .
   \label{zetaknu}
\ee

The results for MA couplnts itself can be obtained putting $\nu=1$. Moreover,
at the point $L_k=0$, i.e. for $Q^2=\Lambda_k^2$, we get ($l=\ln(2\pi)$)
\be
A^{(1)}_{\rm MA}= \frac{1}{2},~~~
\delta^{(2)}_{s}=
-\frac{b_1}{2\pi^2} \, \Bigl(\zeta_1(2)+l\zeta(2)\Bigr),~~
\label{dAmanNew2}
\ee

\section{Integral representations for MA coupling }

As already discussed in Introduction, the MA couplant $A^{(1)}_{\rm MA}(Q^2)$ is constructed as follows:
the LO spectral function is taken directly from PT, and the MA couplant $A^{(1)}_{\rm MA}(Q^2)$ is obtained from the dispersion
integral  (\ref{disp_MA_LO}).

For the $\nu$-derivative of $A^{(1)}_{\rm MA}(Q^2)$, i.e. $\tilde{A}^{(1)}_{\rm MA,\nu}(Q^2)$, there is the following
equation \cite{GCAK}:
\be
\tilde{A}^{(1)}_{\rm MA,\nu}(Q^2)=
\frac{(-1)}{
  \Gamma(\nu)}
\int_{0}^{\infty} \ \frac{d s}{s} r^{(1)}_{\rm pt}(s)
    {\rm Li}_{1-\nu} (-sz)\,,
\label{disptAnuz} 
\ee
where ${\rm Li}_{1-\nu} (-sz)$ is the Polylogarithm presented in Eq. (\ref{Linu}).

At NLO, Eq. (\ref{disptAnuz}) can be extended
in two different ways, which will be shown in following subsections.

{\bf Modification of spectral functions}.~~
The first possibility to extend the result (\ref{disptAnuz}) beyond LO is related to the modification of the spectral function:
\be
\tilde{A}^{(2)}_{{\rm MA},\nu,k}(Q^2) =
\frac{(-1)}{
  \Gamma(\nu)}
\int_{0}^{\infty} \ \frac{d s}{s} r^{(2)}_{\rm pt}(s)
    {\rm Li}_{1-\nu} (-sz_k)\,,
    \label{disptAnuz.ma}
    \ee
where \cite{Nesterenko:2003xb}
\be
r^{(2)}_{\rm pt}(s)=r^{(1)}_{\rm pt}(s)+
\delta^{(2)}_{\rm r}(s) \, 
\label{rima.1}
\ee
and
\be
y=\ln s,~r^{(1)}_{\rm pt}(y)=\frac{1}{y^2+\pi^2} \,,~
\delta^{(2)}_{\rm r}(y)=-\frac{b_1}{(y^2+\pi^2)^2} \, \Bigl[2y f_1(y) +(\pi^2-y^2) f_2(y)\Bigr] \,,
\label{dr5ma}
\ee
with
\be
f_1(y)=\frac{1}{2} \, \ln\bigl(y^2+\pi^2\bigr),~~f_2(y)=\frac{1}{2} -\frac{1}{\pi}\, arctan\left(\frac{y}{\pi}\right) \, .
 \label{f12}
\ee

  For the MA coupling constant itself, we have
   \be
   A^{(i+1)}_{\rm MA,k}(Q^2) \equiv \tilde{A}^{(i+1)}_{{\rm MA},\nu=1,k}(Q^2)    =
   \int_0^{+\infty} \,  
\frac{ d s \, r^{(i+1)}_{\rm pt}(s)}{(s + t_k)}\,.
\label{dispA.ma} 
\ee


{\bf Modification of Polylogaritms}.~~
The NLO results (\ref{disptAnuz}) can also be expanded with the $\hat{R}_1$ operators shown in (\ref{hR_i}), and this results in the following result:
\be
\tilde{A}^{(2)}_{{\rm MA},\nu,1}(Q^2) =
\int_{0}^{\infty}  \frac{d s}{s} r^{(1)}_{\rm pt}(s)
\tilde{\Delta}^{(2)}_{\nu,1}(sz_1)\,,
    \label{disptAnuz.ma1}
    \ee
    where the results for $\tilde{\Delta}^{(2)}_{\nu,1}(z)$ can be found in
    (\ref{Lii.1}).

    The results for MA coupling constant itself can be obtained from (\ref{disptAnuz.ma1}) putting $\nu=1$.

\section{Conclusions}

In this short paper, we have demonstrated the results obtained in our recent paper \cite{Kotikov:2022sos} (see also
\cite{Kotikov:2022vnx}). In particular,  \cite{Kotikov:2022sos} contains $1/L$-expansions of $\nu$-derivatives of
the strong couplant $a_s$ expressed as combinations of the operators $\hat{R}_i$ (\ref{hR_i}) applied to the LO couplant $a_s^{(1)}$.
Using the same operators to $\nu$-derivatives of LO MA couplant $A_{\rm MA}^{(1)}$, four different
representations were obtained for $\nu$-derivatives of MA couplant,
i.e. $\tilde{A}_{\rm MA,\nu}^{(i)}$, in each $i$-order of PT.
All results are presented in \cite{Kotikov:2022sos,KoZe23} up to the 5th order of PT, where the corresponding
coefficients of  QCD $\beta$-function  are well known
(see \cite{Baikov:2016tgj}). In this paper, we have limited ourselves to the first two orders in order to exclude the most cumbersome
results obtained for the last three PT orders.

In the case of MA couplant,
high-order corrections are negligible in both asymptotics: $Q^2 \to 0$ and $Q^2 \to \infty$, and are nonzero in a neighborhood of
the point $Q^2 =\Lambda^2$.
Thus, in fact, they  represent only minor corrections to LO MA couplant $A_{\rm MA}^{(1)}(Q^2)$.\\


{\bf Acknowledgments}~
This work was supported in part by the Foundation for the Advancement
of Theoretical Physics and Mathematics “BASIS”.
One of us (A.V.K.) thanks the Organizing Committee of the
XXVth International Baldin Seminar on High Energy Physics Problems
Relativistic Nuclear Physics and Quantum Chromodynamics
(September 18-23, Dubna, Russia)
for invitation.

\end{document}